\documentclass[12pt,preprint]{cmlpasp}
\usepackage{multirow}
 \newcommand{\beq}{\begin{equation}}
 \newcommand{\eeq}{\end{equation}}
 \newcommand{\beqn}{\begin{eqnarray}}
 \newcommand{\eeqn}{\end{eqnarray}}

\begin{document}
\title{Detecting Exoplanet Transits
through Machine Learning Techniques with Convolutional Neural Networks
}
\author{Pattana Chintarungruangchai and Ing-Guey Jiang}

\affil{
{Department of Physics and Institute of Astronomy,}\\
{National Tsing-Hua University, Hsin-Chu, Taiwan} 
}
\email{jiang@phys.nthu.edu.tw}

\begin{abstract} 
A machine learning technique 
with two-dimension convolutional neural network
is proposed for detecting exoplanet transits. 
To test this new method, 
five different types of deep learning models with or without folding 
are constructed and studied. The light curves of 
the Kepler Data Release 25
are employed as the input of these models.
The accuracy, reliability, and completeness are determined
and their performances are compared. 
These results indicate that 
a combination of two-dimension convolutional neural network
with folding would be an excellent choice for 
the future transit analysis.   
\end{abstract}

\newpage

\section{Introduction}
\label{sec:intro}

The transit method has successfully detected many new extra-solar planets
(exoplanets) with the Kepler Space Telescope being a major player 
(Koch et al. 2010).  The success of Kepler mission depended in
part on the Kepler pipeline (Jenkins et al. 2010) 
to process the data, identify
transit signals and validate their existence. 

On the other hand, 
during recent years, machine learning becomes widely used in many fields, 
including astronomy (Cuevas-Tello, Tino, \& Raychaudhury 2006). 
There are many types of machine learning techniques, 
and some of them have already been used to analyse transit light curves. 
For example, k-nearest neighbors (kNN) by Thompson et al. (2015), 
decision tree by Coughlin et al. (2016), 
random forest by McCauliff et al. (2015), 
and self-organizing map (SOM) by Armstrong et al. (2017). 
 
In addition, deep learning, a powerful machine learning technique,  
was also used to analyse light curves.
Shallue \& Vanderburg (2018) 
employed deep learning to vet the signals
which were first detected by the Box-Least-Square algorithm 
(Kovacs et al. 2002) and successfully discovered new transits
from Kepler light curves.
Moreover, without Box-Least-Square algorithm, 
Pearson et al. (2018) suggested to directly use
deep learning to examine the folded Kepler light curves
to search for transits. Please note that the folding 
was one of the main steps in Box-Least-Square algorithm.

Their results showed that deep learning can play an  
important role for the detection of exoplanets.
However, in these previous work, only one-dimension  
convolutional neural network (1D-CNN) was used.
The input of an 1D-CNN model is 
a one-dimension array which has the flux values at different
time of a light curve.  
When folding is used, 
one needs to add up all folded light curves and take averages 
for flux values. The signals of transits can be enhanced
only when the folding period is exactly the same as the transit period.  
When one searches for new transits with unknown transit periods,
the resolution of trial periods which are employed as folding periods
needs to be very high.
To solve this problem,  
the method of 
two-dimension convolutional neural network (2D-CNN),
which was mainly used for pattern recognition, is explored 
here. All flux values of folded light curves can be kept
and the transit signals would not be averaged out even 
when the folding period is different from the transit period.  
In this paper, both 1D-CNN and 2D-CNN deep learning models with 
phase folding will be constructed.
Based on convolutional neural network (CNN), we 
study several deep learning models and compare their performances. 

In Section \ref{sec:ml}, 
the basic concept of machine learning,
deep learning, and CNN will be introduced. 
In Section \ref{sec:model}, five deep learning models are constructed 
based on convolutional neural network (CNN).
The samples of light curves used as training, validation, and testing 
will be described in Section \ref{sec:laica}.
The results and the demonstration will be presented in 
Section \ref{sec:compa} and Section 6.
Conclusions would be made in Section \ref{sec:disca}.

\section{Artificial Intelligence}
\label{sec:ml}

Artificial Intelligence (AI) is a computer program that can do some 
tasks automatically.
For example, when we try to investigate whether there are planets 
moving around stars or not, we can give observed light-curve data 
of stars to an AI program.  After the AI does some analysis, 
it will give an answer that there are planets or not around a particular star.
Depending on the design of this AI program, in addition to showing 
the existence of planets, more parameters such as orbital periods
could be listed as part of the output.

In order to make it clear how an AI program can be constructed,
in this section, the concepts of machine learning, deep learning, and 
neural network will be introduced.
The structures and equations 
of convolutional neural network will also be described.

\subsection{Machine Learning}
\label{ssec:ml}
 
It should be convenient if a computer program can learn to program itself. 
Machine learning is one type of AI algorithm that makes computer programs 
try to learn something and change themselves, 
in order to improve the analysis and give correct results.
Generally, the training data are employed
to make computer programs learn and recognize 
the relation between the input and output.   
Some machine learning techniques, such as kNN, 
would memorize the training data and use it for later analysis. 
In this case, ``learning'' means memorizing data. 
This type of techniques is called non-parametric algorithm.

On the other hand, most machine learning techniques
will not memorize the training data, but will
keep changing the values of some parameters during the learning process.
It is called parametric algorithm for this case.
 
\subsection{Logistic Regression}
\label{ssec:lori}

One example of parametric algorithm is logistic regression, 
which is a very simple machine learning technique that often 
used in data classification. 
For example,  
if we want to classify light curves into two classes, 
with or without planet transits, 
the logistic regression is a linear equation as
\begin{equation}
\label{equ:lori}
h(\mathbf{x}) = \sum_{i=1}^{n}{x^{(i)}w^{(i)}} + b =  x^{(1)}w^{(1)}+x^{(2)}w^{(2)}+...+x^{(n)}w^{(n)} + b,
\end{equation}
where the input data $x^{(i)}$ is the light-curve flux, 
$w^{(i)}$ and $b$ are parameters, $n$ is the number of input data. 
Thus, there are $n+1$ parameters in this logistic regression model. 
The output of this model depends on $h(\mathbf{x})$ as
\begin{equation} \label{equ:lori_y}
 y=\left\{ 
\begin{array}{ll}
0 & {\rm when}\  h(\mathbf{x}) < 0;\\
1 & {\rm when}\  h(\mathbf{x}) \ge 0, 
\end{array} \right. 
\end{equation}
where $y$ is the classification result that could be 0
(no transit) or 1 (with transit). 
 
\begin{figure}[ht]
\centering
\includegraphics[width=0.8\textwidth]{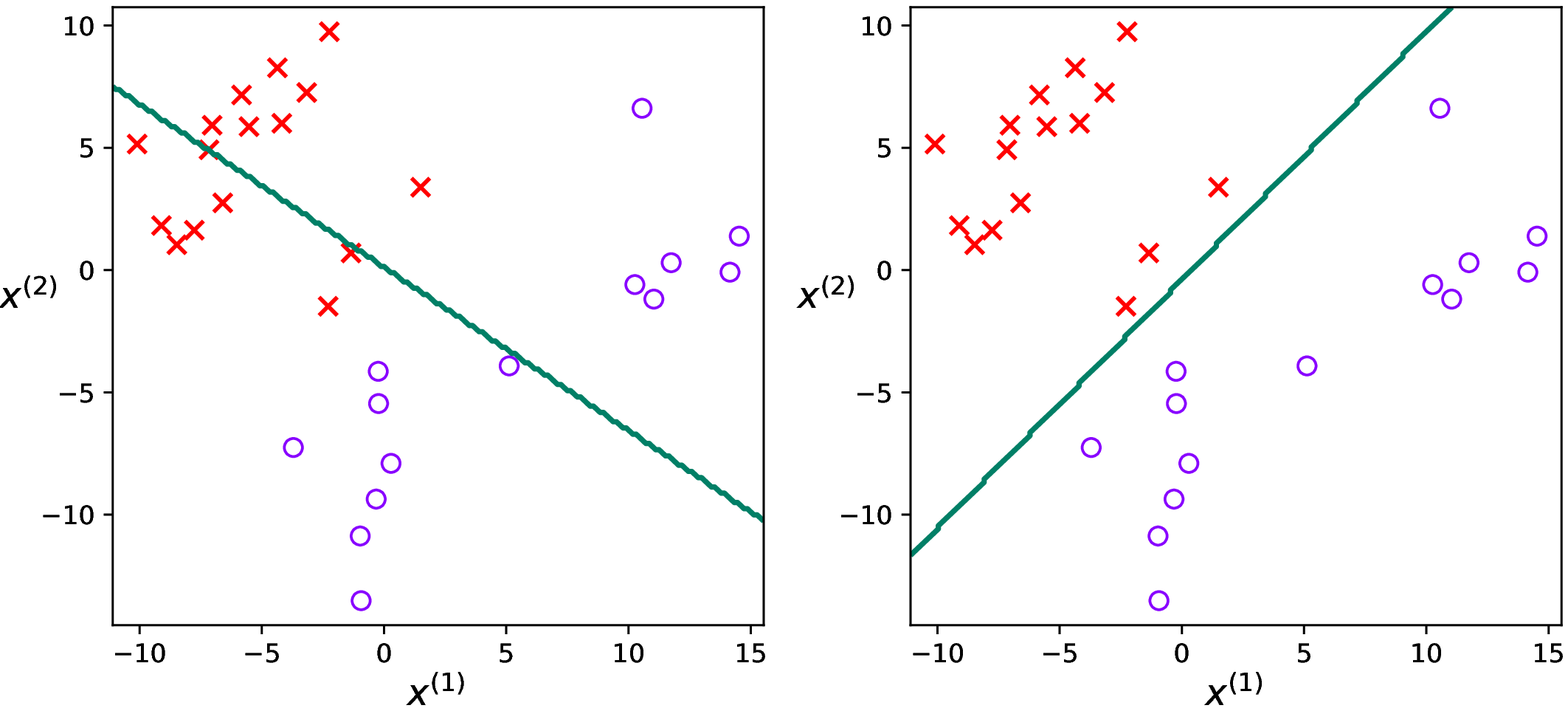}
\caption{The classification by logistic regression with unlearned model (left)
 and learned model (right). Circles and crosses indicate different group
 of data. The straight line is the data separator.} 
\label{fig:lori}
\end{figure}

To give a simple example, we assume that the output $y$ only depends on  
$x^{(1)}$ and $x^{(2)}$, the logistic regression
will include three parameters $w^{(1)}$, $w^{(2)}$ and $b$. 
A possible learning process is presented in Fig. \ref{fig:lori}. 
In the plots, each point gives the true classification:
crosses are the cases with transits and circles are those without transits. 
The left panel shows the unlearned model that 
the values of $w^{(1)}$, $w^{(2)}$ and $b$ are randomly given. 
The solid line is the border of two classifications
determined from Eq. (\ref {equ:lori})-(\ref {equ:lori_y}). 
It is clear that this line does not separate crosses and circles well.
The right panel is the learned model with new values
of $w^{(1)}$, $w^{(2)}$ and $b$. The corresponding solid line
does separate crosses and circles well. From the values of 
$x^{(1)}$ and $x^{(2)}$, the model can give a correct answer
whether there is transit or not.
Thus, the ``learning'' is to change the values of parameters 
in order to obtain the correct answer. 

\subsection{Deep Learning}
\label{ssec:ann}

For most machine learning techniques, we have to choose some 
appropriate features of input data that might influence the output answer. 
For example, for the light-curve analysis
we may calculate and extract some features such as
the mean, the standard deviation, 
the period, the amplitude, or the 
auto-correlation function (Kim et al. 2011). 
The process of selecting features is called feature engineering. 
However, there are some algorithms that can make programs learn to 
find features. They are called representation learning. 
One of the well known representation learning techniques 
is deep learning.

\begin{figure}[ht]
\centering
\includegraphics[width=0.7\textwidth]{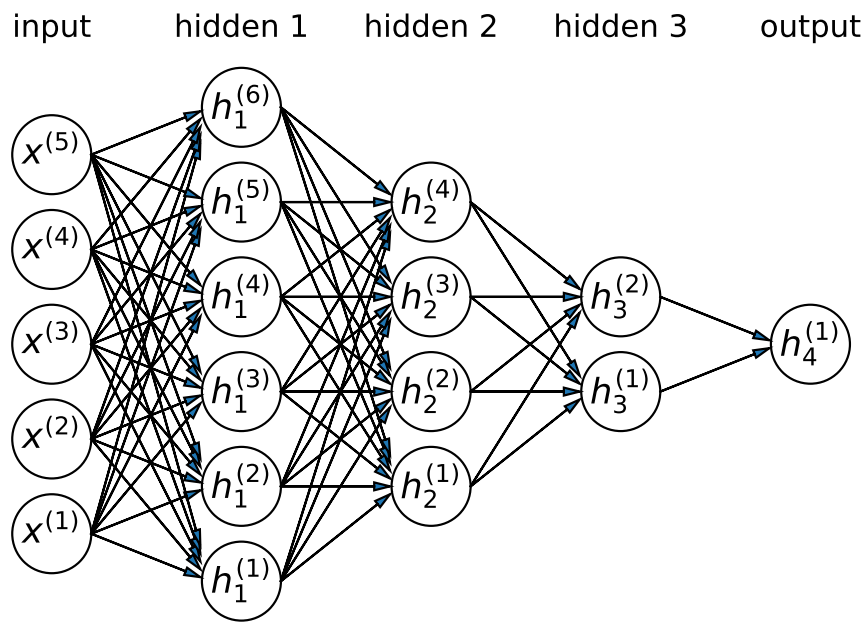}
\caption{The structure of ANN. Each column of circles is one layer. 
The data will flow through all layers from left to right.} 
\label{fig:ann}
\end{figure}

Deep learning is one type of machine learning algorithms which is proven 
to be the most powerful one these days. It can extract some features 
itself. The core of deep learning is a structure 
called artificial neural network (ANN). The concept of ANN 
imitates the human brain. It consists of many calculation units 
called neuron. There are various designs of ANN. A simplest type of ANN is 
multilayer perception (MLP), which consists of many layers called 
fully connected layers. It usually consists of input layer, 
hidden layers, and output layer. 
A number of neurons are assigned to be within each layer.  
Neurons would accept input data, do 
some calculations, and give output. 
The structure of an ANN is shown in Fig. \ref{fig:ann} as an example.
There could be one or many hidden layers. 

The input data flows into 
neurons of the input layer, then give their output as the input of 
neurons of the first hidden layer. 
The data goes through layers one by one.    
Finally, the data will flow from the last hidden layer 
into the output layer. Then, we obtain the answer of the problem
in the end. 

The calculation in each neuron is a linear function, 
followed by a nonlinear function, 
i.e. activation function $\phi$, as
\begin{eqnarray} \label{equ:ann}
h_1^{(j)} &=& \phi(\sum_{i=1}^{m_1}{x^{(i)}w_1^{(i,j)}} +b_1^{(j)}) \nonumber \\
h_2^{(j)} &=& \phi(\sum_{i=1}^{m_2}{h_1^{(i)}w_2^{(i,j)}} +b_2^{(j)}) \nonumber \\
...&&...  \nonumber \\
h_{n-1}^{(j)} &=& \phi(\sum_{i=1}^{m_{n-1}}{h_{n-2}^{(i)}w_{n-1} ^{(i,j)}} +b_{n-1}^{(j)} ) 
\nonumber \\
h_n &=& \sum_{i=1}^{m_n}{h_{n-1}^{(i)}w_n^{(i)}} +b_n 
\end{eqnarray}
and
\begin{equation} \label{equ:ann_y}
 y=\left\{ 
\begin{array}{ll}
0 & {\rm when}\ h_n < 0;\\
1 & {\rm when}\ h_n \ge 0. 
\end{array} \right. 
\end{equation} 
Here $(i)$ indicates features of input in each layer, 
$(j)$ indicates features of output, and
$h_1^{(j)}$,$h_2^{(j)}$,... are the output of layers. 
There are many types of activation functions. 
Usually, rectified linear unit function (ReLU)  
$\phi(x)=max\{0,x\}$ is used (Nair \& Hinton 2010). 
Note that the output layer $h_n$ does not use any activation function,
and there is no index $(j)$ for $h_n$ because here 
we only need one output. The parameters
$w_1^{(i,j)}$,$w_2^{(i,j)}$,... and $b_1^{(j)}$,$b_2^{(j)}$,... 
in layers will be changed during the learning process. 
The important features can be picked up gradually through the 
updated values of these parameters.
We do not have to do feature engineering. 

\subsection{Convolutional Neural Network}
\label{ssec:cnn}

\begin{figure}[ht]
\centering
\includegraphics[width=0.7\textwidth]{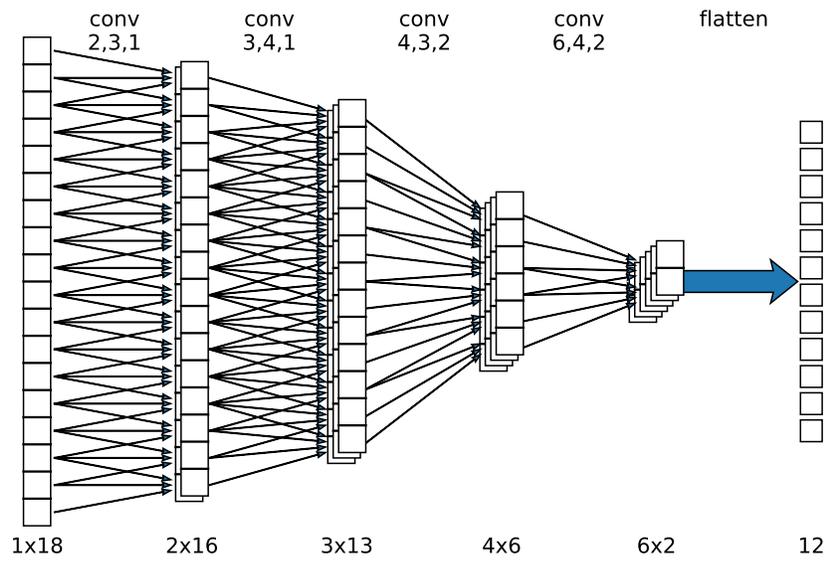}
\caption{An example of the structure of convolutional layers of 1D-CNN. 
The number below is channel number$\times$data number in each layer. 
The numbers over are the output channel number, kernel size, and stride, 
respectively.} 
\label{fig:cnn}
\end{figure}

The convolutional neural network (CNN) is an alternative of MLP. 
In plain MLP, each data point is treated as an individual feature, 
and the spatial structure of data points is ignored. 
In contrast, CNN can take into account the spatial structure, which is 
very important for time series data or picture analysis.
CNN includes convolutional layers and fully connected layers. 
The input and output of each convolutional layer of CNN may have 
many channels. Each channel consists of continuous data points.
In a convolutional layer, the input data points of channels 
would do the convolution with 
a filter called kernel. The calculations in CNN are as
\begin{eqnarray} \label{equ:cnn1d}
h_1^{(j,u)} &=&
\phi\left(\sum_{i=1}^{c_0}\sum_{l=1}^{k_1}w_1^{(i,j,l)}x^{(i,s_1(u-1)+l)}+b_
1^{(j)}\right) \nonumber  \\
h_2^{(j,u)} &=&
\phi\left(\sum_{i=1}^{c_1}\sum_{l=1}^{k_2}w_2^{(i,j,l)}h_1^{(i,s_2(u-1)+l)}
+b_2^{(j)}\right) \nonumber \\
...&&... \nonumber \\
h_n^{(j,u)} &=&
\phi\left(\sum_{i=1}^{c_{n-1}}\sum_{l=1}^{k_n}w_n^{(i,j,l)}h_{n-1}^{(i,s_n(u
-1)+l)}+b_n^{(j)}\right), 
\end{eqnarray}
where $i$ and $j$ are the channel index of input and output, 
$u$ is the position index of output, 
$l$ is the position index in kernel.
Thus, 
$w_n^{(i,j,l)}$ is the weight parameter of 
point $l$ of kernel for input channel $i$ to output channel $j$. 
In addition, $k_n$ is the kernel size, 
$c_0$ is the input channel number of first layer, 
$c_{n-1}$ is the output channel number of ($n-1$)th layer 
and also the input channel number of $n$th layer. 
After passing one convolutional layer, 
the output size of data in each channel becomes
\begin{equation}
\mu_n = \frac{\mu_{n-1}-k_n}{s_n}+1
\end{equation}
where $s_n$ is the stride. A stride is the step size of skipping  
after calculating one output value. 
After passing through all convolutional layers, 
the outcome value will be flatten into separated points
\begin{equation} \label{equ:flat}
X^{((i-1)\mu_n+u)} = h_n^{(i,u)},
\end{equation}
where $\mu_n$ is the output size of $n$th layer. 
After flattening, $X$ becomes the input of a fully connected layer.
An example of CNN that input data 
is a light curve with 18 points which pass through four convolutional layers 
is shown in Fig. \ref{fig:cnn}. The numbers over each layer are 
the output channel number, kernel size, and stride, respectively.

CNN can be one or two dimensional. 
1D-CNN is usually used for time series data and 
2D-CNN is often used in picture analysis.
Here we use 2D-CNN for light-curve data as the first time in this field.
The calculations of 2D-CNN are
\begin{eqnarray} \label{equ:cnn2d}
h_1^{(j,u,v)} &=&
\phi\left(\sum_{i=1}^{c_0}\sum_{l=1}^{k_{1,1}}\sum_{m=1}^{k_{2,1}}w_1^{(i,j,
l,m)}x^{(i,s_{1,1}(u-1)+l,s_{2,1}(v-1)+m)}+b_1^{(j)}\right) \nonumber \\
h_2^{(j,u,v)} &=&
\phi\left(\sum_{i=1}^{c_1}\sum_{l=1}^{k_{1,2}}\sum_{m=1}^{k_{2,2}}w_2^{(i,j,
l,m)}h_1^{(i,s_{1,2}(u-1)+l,s_{2,2}(v-1)+m)}+b_2^{(j)}\right) \nonumber \\
...&&... \nonumber \\
h_n^{(j,u,v)} &=&
\phi\left(\sum_{i=1}^{c_{n-1}}\sum_{l=1}^{k_{1,n}}\sum_{m=1}^{k_{2,n}}w_n^{(
i,j,l,m)}h_{n-1}^{(i,s_{1,n}(u-1)+l,s_{2,n}(v-1)+m)}+b_n^{(j)}\right),
\end{eqnarray}
where $u$ and $v$ are the position index of output dimension 1 and dimension 2, 
$k_{1,n}\times k_{2,n}$ is the kernel size.

\subsection{Stochastic Gradient Descent}
\label{ssec:sgd}

For a logistic regression or deep learning model,
the method called stochastic gradient descent (SGD)
is usually used to decide how to change parameters 
during the learning process. 
This method is to compare the answer that predicted from the model 
with the true answer of a training data set, 
and calculate the cost function which indicates the badness of 
the predicted answer. 
For classification problems, the most popular
cost function is the cross entropy   
\begin{equation} \label{equ:sgd}
C(\mathbf{z}) = -\frac{1}{M}\sum_{i=1}^{M}(y_i\log y'_i+(1-y_i)\log(1-y'_i)),
\end{equation}
where $y_i$ is the predicted answer 
from one particular training data set, 
$y'_i$ is the true answer of the same data set, 
$M$ is the number of training data sets 
used in one turn of parameter updating calculation,
and $\mathbf{z}$ means all parameters such as 
$w_1^{(i,j)}, b_1^{(j)}, b_2^{(j)}$ etc. in the model.
 
During the training process, Eq. (\ref{equ:ann_y}) is replaced by
a smooth function
\begin{equation} \label{equ:sigmoid}
y=\frac{1}{1+e^{-h_n}}.
\end{equation}
It is a sigmoid function such that $y$ can be any value 
ranging from 0 to 1. 
It represents the probability of detecting a transit. 
Therefore, the cost function is a smooth function 
with respect to any parameter $z_i$ in the model.
 
The objective is to determine $\mathbf{z}$ which would minimize 
$C(\mathbf{z})$.
One method is to calculate 
\begin{equation} \label{equ:g_i}
g_i = \frac{\partial C(\mathbf{z})}{\partial z_i}
\end{equation}
and change the values of parameters by
\begin{equation} \label{equ:change}
z_{i,t} = z_{i,t-1}-\eta g_{i,t-1},
\end{equation}
where $\eta$ is the learning rate, 
$t$ is the index of one turn of parameter updating.
Thus, $t=1,2,3, ...$ until the training process is completed.
    
There are many alternative forms of SGD, and 
Adam optimization algorithm (Kingma \& Ba 2015) is employed in this paper. 
During the training process, parameters are updated through 
\begin{equation} \label{equ:adam}
z_{i,t} = z_{i,t-1}-\eta \frac{m_t}{1-\beta_1^t}
\left(\sqrt{\frac{v_t}{1-\beta_2^t}}+\epsilon\right)^{-1},
\end{equation}
where $\beta_1$, $\beta_2$ and $\epsilon$ are Adam's parameters, and
\begin{eqnarray} \label{equ:adammv}
m_t &=& m_{t-1} + (1-\beta_1)g_{i,t-1} \nonumber  \\
v_t &=& v_{t-1} + (1-\beta_2)g_{i,t-1}^2,
\end{eqnarray}
where $m_0=v_0=0$. 
We use the above Adam optimization with 
$\eta=0.001$, $\beta_1=0.9$, $\beta_2=0.999$, $\epsilon=10^{-8}$. 

\section{The Models}
\label{sec:model}

For given noisy light curves, it could be difficult 
for conventional approaches
to directly confirm whether there is any transit or not.
Later validation and vetting are needed.
Deep learning models can be designed to do the task of transit detection or 
vetting, depending on the input data and how they are trained. 
In this paper, we study possible better deep learning models
on direct transit detection.

Let us review the most recent related work here.
Zucker \& Giryes (2018) proposed that 1D-CNN can be useful
for the detection of exoplanet transits. 
They generated hypothetical light curves with modeled 
red noise. These artificial data are then used to train and test the code.
They recommended 1D-CNN as an excellent method for future transit analysis  
of huge amount of space mission data.
To take the advantage of transit periodicity, 
Pearson et al. (2018) used the phase folding technique. 
After the folding, the transit signal will be enhanced,
and the transit detection  efficiency could be increased.
The transit period has to be known in advance for the  
phase folding technique. This is not a problem 
for training data sets but it is a difficulty 
when we search transits from future observational light curves.

\begin{figure}[ht]
\centering
\includegraphics[width=0.7\textwidth]{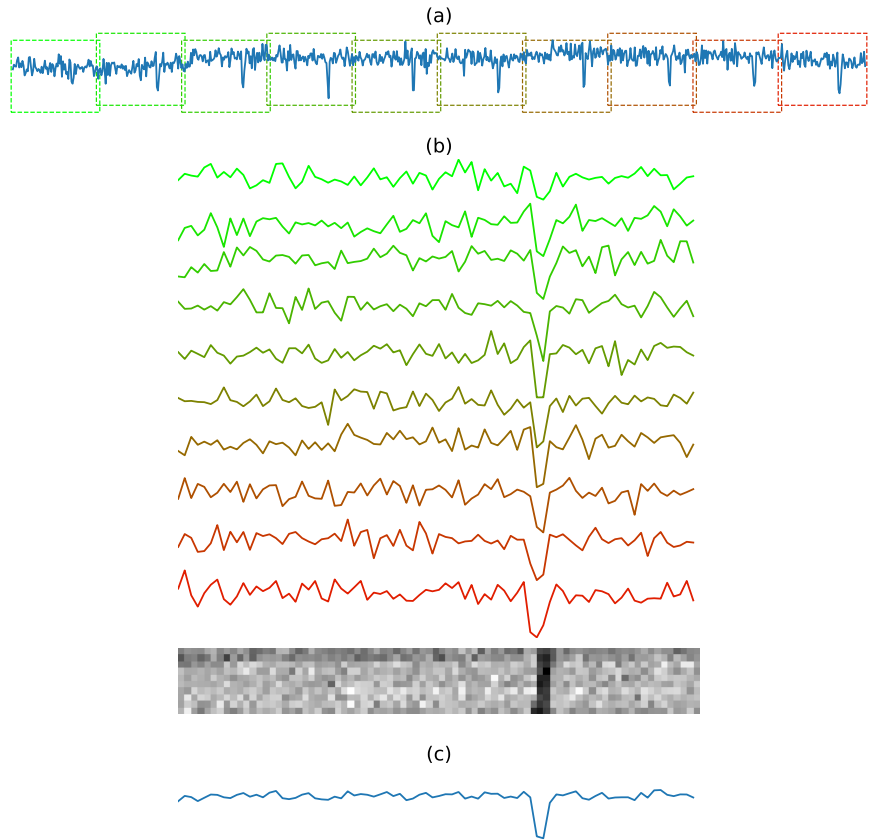}
\caption{The folding process 
when the folding period is the same as the transit period.
(a) The light curves before folding. 
(b) The light curves after folding. 
(c) The mean of periods after folding.} 
\label{fig:fefo1}
\end{figure}

\begin{figure}[ht]
\centering
\includegraphics[width=0.7\textwidth]{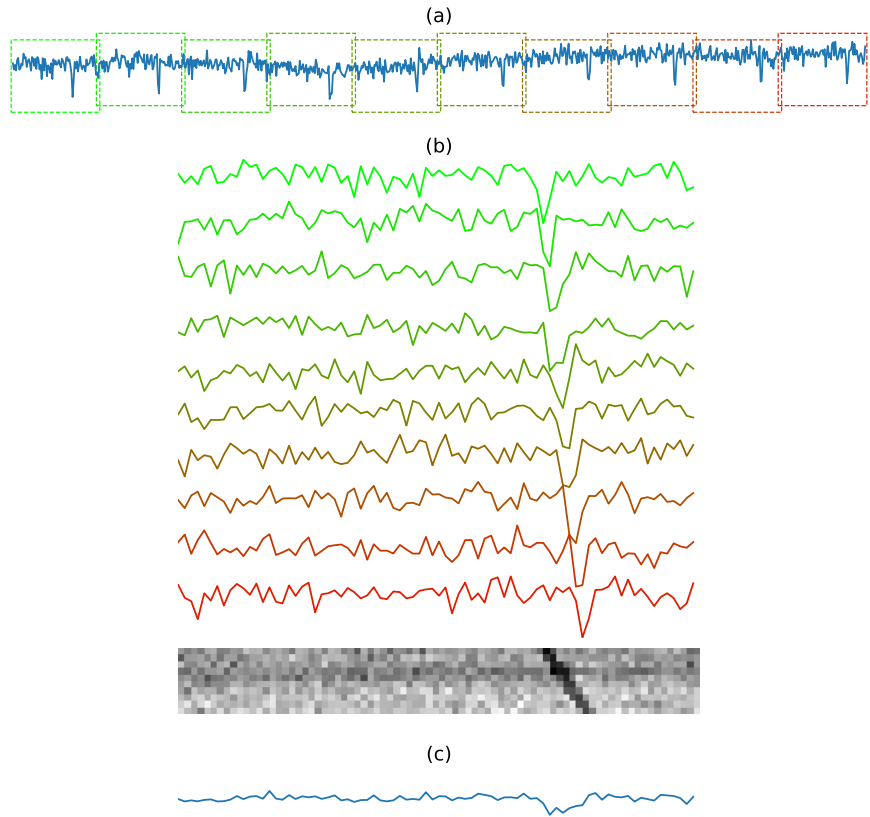}
\caption{The folding process 
when the folding period is different from the transit period.
(a) The light curves before folding. 
(b) The light curves after folding. 
(c) The mean of periods after folding.
} 
\label{fig:fefo2}
\end{figure}

Therefore, in order to address this issue, here 
we construct 2D-CNN models.
Fig. \ref{fig:fefo1} presents the input of a model of 2D-CNN 
with folding, where (a) 
shows ten transits; (b) shows the folding with the same period;
(c) shows the mean of all folded light curves.  
The method of Pearson et al. (2018) would use (c) as input data,
which has a very clear transit signal. The 2D structure of 
folded light curves is shown at the bottom of (b) and the vertical dark band
is the transit signal.
A 2D-CNN model will use (b) as input data, 
and it is likely to get the same answer as the 1D-CNN model 
of Pearson et al. (2018). 
Fig. \ref{fig:fefo2} demonstrates
the advantage of 2D-CNN with folding, where (a)
shows the input of ten transits; (b) shows the folding with 
different period; (c) shows the mean of all folded light curves.
Since the folding period is 
different from the transit period, the flux drop looks as 
keep shifting to slightly different phase.
Thus, the mean transit signal in (c) becomes unclear
and the 1D-CNN model of Pearson et al. (2018) 
might fail to give a correct answer about transit detection.
The 2D structure of 
folded light curves is also shown at the bottom of (b) and the dark band
is the transit signal. The tilted dark band is clear, 
so a model of 2D-CNN with folding shall be able to detect 
this transit successfully.

\begin{figure}[ht]
\centering
\includegraphics[width=0.7\textwidth]{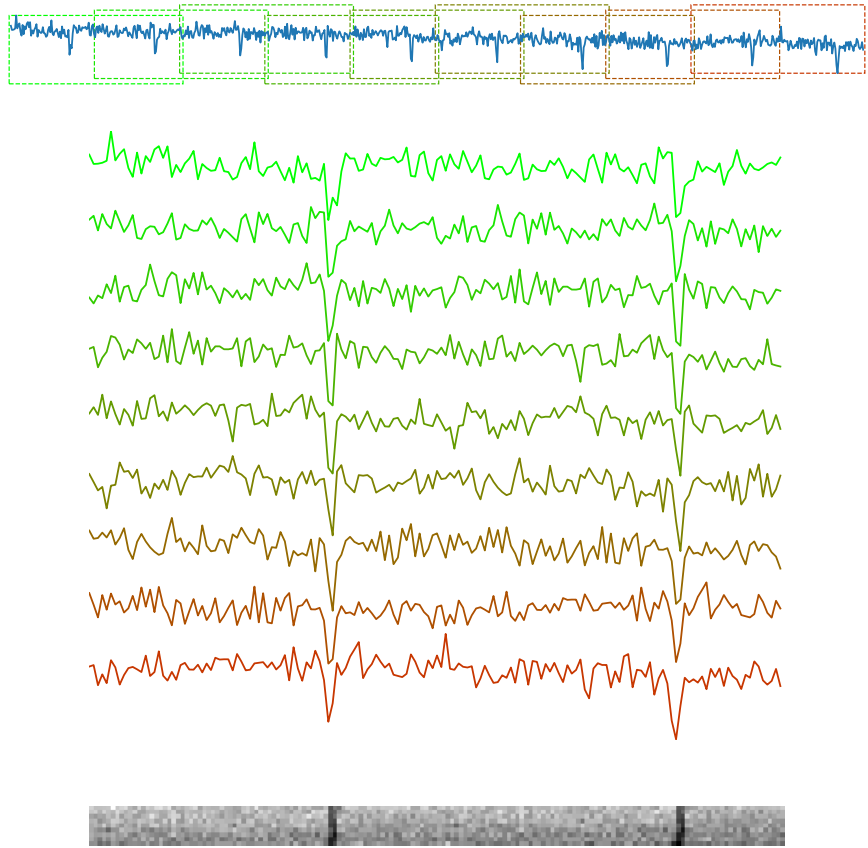}
\caption{The folding process 
when the folding period is two times of the transit period.
Only the first and last transit periods are used one time, 
all the rest are used for two times.
(a) The light curves before folding. 
(b) The light curves after folding. 
(c) The mean of periods after folding.
} 
\label{fig:fefo3}
\end{figure}

Moreover, we also invent a new way of folding 
which uses two transit periods in one folding as shown in 
Fig. \ref{fig:fefo3}. 
Every transit period, except the first and the last, is used for two times. 
We call this model as 2D-CNN-folding-2 in this paper.

In order to understand the advantages of our 2D-CNN models, we 
construct five deep learning models as listed 
in Table \ref{tab:model}.

\begin{table}[h]
\begin{center}
\begin{tabular}{llll}
\hline
Model & Type & Input &Convolutional Layers 
[channel number, kernel size, stride]  \\ \hline
1 & MLP & 4000 & - \\
2 & 1D-CNN-folding-0 & 1$\times$4000 & 
[40,40,4],[32,23,4],[32,23,4],[32,24,4] \\
3 & 1D-CNN-folding-1 & 1$\times$400 & [16,8,2],[32,5,2],[64,5,2],[64,5,2] \\
4 & 2D-CNN-folding-1 & 1$\times$10$\times$400 & 
[16,3$\times$8,1$\times$2],[32,3$\times$5,1$\times$2],[64,3$\times$5,1$\times$2],[64,3$\times$5,1$\times$2]\\
5 & 2D-CNN-folding-2 & 1$\times$9$\times$800 &[16,3$\times$16,1$\times$4],[32,3$\times$5,1$\times$2],[64,3$\times$5,1$\times$2],[64,3$\times$5,1$\times$2]\\
\hline
\end{tabular}
\caption[Table R]{Five deep learning models.}
\label{tab:model}
\end{center}
\end{table}
Model 1 is just a MLP with three fully connected layers. 
Model 2-5 are CNN models with four convolutional layers. 
The channel number, kernel size, and stride of each layer
are shown in Table \ref{tab:model}. 
For 2D-CNN models, the kernel size and stride are 2D, shown as $k_1\times k_2$. 
Furthermore, after convolutional layers, all models consist of three 
fully connected layers as in Model 1. 
The output sizes of these three layers are 256, 128, 1, respectively.

Model 2, called 1D-CNN-folding-0, is a 1D-CNN model that 
uses whole light curves without folding. 
Previous work such as Zucker \& Giryes (2018) and Hinners et al. (2018) 
employed models like this. 
The basic structures of our convolutional layers followed 
Zucker \& Giryes (2018), but the input data number is different, 
and we use three fully connected layers.
Model 3, called 1D-CNN-folding-1, is 1D-CNN with folding 
as in Pearson et al. (2018).
The mean of folded light curves are calculated
and used as the input data. 

\begin{figure}[ht]
\centering
\includegraphics[width=0.3\textwidth]{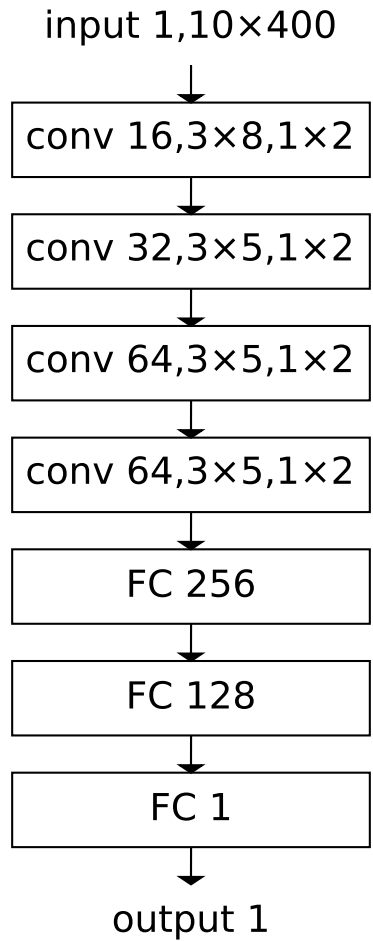}
\caption{The structure of Model 4. 
The numbers after "input" are the channel number 
and the data size (height*width). 
The convolution layer is indicated as "conv" 
and the numbers are output channel number, kernel size 
and stride, respectively. 
The fully connected layer is indicated as "FC"  
and the number is the output size.} 
\label{fig:cnn2d}
\end{figure}

Model 4, called 2D-CNN-folding-1, is a 2D-CNN model with folding. 
The kernel size and stride of the second dimension 
are identical to that of Model 3. 
The structure of this model is shown in Fig. \ref{fig:cnn2d}.
Model 5, called 2D-CNN-folding-2, is an alternative of Model 4. 
All details are the same as Model 4 except that it 
uses two periods of data for each folding.

To build up the computer programs for the above five models,
the software Pytorch (Paszke et al. 2017) 
is employed.  Through Pytorch, deep learning models
can be constructed from elements of tensors and functions.

\section{Light Curve Data}
\label{sec:laica}

The observations of Kepler Space Telescope started from 2nd May 2009 
and ended on 11th May 2013. The data is divided 
into 18 quarters,  i.e Q0 - Q17. 
Each quarter's length is about three months, though some quarters 
are shorter. The data include long cadence which took images 
every 30 minutes, and short cadence which took images 
every 2 minutes. In this work, only long cadence data
Q1 - Q17 is considered,
as there are not that many stars observed with short cadence. 
Through Mikulski Archive for Space Telescopes (MAST), 
Data Release 25 (DR25) were downloaded
(https://doi.org/10.17909/T9488N). 
  
The light curve data consist of many columns including two types of the 
observed fluxes. One is simple aperture photometry (SAP), 
which is the flux obtained by direct photometric analysis 
and may still include some artifact. 
Another is pre-search data conditioning (PDC), 
which the artifact was already removed (Smith et al. 2012).
We use PDC data.
However, the outliers from astrophysical events, 
such as stellar flares and microlensing events, are not removed 
from PDC light curves. We therefore remove
data points that are 6$\sigma$ higher than interpolated curves.

Light curves from more than 200,000 stars have been analyzed by 
the Kepler team (Borucki et al. 2009, Batalha et al. 2013, 
Coughlin et al. 2016, Christiansen et al. 2013, 2015, 
Mullally et al. 2015, 2016, 
and Thompson et al. 2018) through the Kepler pipeline. 
As they described, Transit Planet Search (TPS) module detrends 
the light curves and identifies possible signals 
called threshold crossing events (TCEs).
Data Validation (DV)
checks that the signals are likely real and creates reports.  
The Robovetter (Thompson et al. 2018) then vets the signals and 
creates a catalog of Kepler object of interest (KOI).
Those data which are confirmed to be nothing
to do with planet transits will be flagged as false positive. 
The rest are called candidates.

The KOI catalog includes 4717 planetary candidates from 3607 stars, 
and 2299 of them have been confirmed as planets.
The Exoplanet Archive, which is one of NASA Archives,
provides the list of confirmed transit planets 
and the properties of planets and central stars, for example, 
orbital periods, transit depths, impact parameters.

On the other hand, we also need those non-transit light curves to be
training data sets. In fact, the Kepler team failed to find transits
among most of the light curves from more than 200,000 stars,
and these can be defined as ``Kepler non-transit light curves''.
Although it is possible that there could be hidden transit signals 
caused by planets smaller than those discovered ones  
in Kepler non-transit light curves, we decide to ignore them 
in this paper.
That is, smaller planets are considered to be out of the scope 
of this paper, and their possible signals are treated as 
part of noise in Kepler non-transit light curves.

In order to train our models, we need to have 
a half of samples to be transit light curves, 
and another half to be non-transit light curves. 
Considering synthetic planets with the same size
distribution and size range as discovered ones,
artificial transit light curves are generated by 
adding the flux drop of transit signals on 
Kepler non-transit light curves.
The non-transit light curves are simply defined to be 
Kepler non-transit light curves.
We then can have equal number of transit and non-transit light curves
to train our deep learning models.
  
The transit signals are modeled as in Mandel \& Agol (2002). 
The related parameters include the transit period ($\tau$), 
the ratio of orbital semi-major axis 
to stellar radius ($a/r_s$),  the ratio of 
planet radius to stellar radius ($r_p/r_s$), 
the inclination ($i$), the 
linear limb darkening coefficient ($u_1$), 
and the quadratic limb darkening coefficient ($u_2$). 
The flux drop is proportional to $(r_p/r_s)^2$, so
the signal-to-noise ratio is defined to be 
a parameter SNR as
\begin{equation} \label{equ:snr}
{\rm SNR} = \frac{(r_p/r_s)^2}{\sigma_{lc}}\sqrt{n},
\end{equation}
where $\sigma_{lc}$ is the standard deviation of the flux values 
of a given normalized light curve and $n$ is the number 
of observed points within the transit duration. 
To generate transit light curves, we need to set values of these
parameters. The orbital period, 
i.e. transit period, $\tau$ is set to range from 0.84 
to 8.4 days and distribute uniformly in ${\rm log}(\tau)$ space.
The parameter $a/r_s$ is derived from $\tau$.
The ratio $r_p/r_s$ follows the same 
distribution and range as discovered ones in the Exoplanet Archive.
S/N is then determined and we exclude the cases with S/N < 7.
The inclination $i$ is uniformly distributed from 85 to 90 degree.
The coefficients $u_1$ and $u_2$ are set to have 
normal distributions. The mean and standard deviation
are assumed to be the same as the ones 
of $u_1$ and $u_2$ of Kepler transit data.
We also exclude the cases which make transit duration shorter than 0.04 day, 
i.e. two times of the observing cadence of Kepler long cadence data. 
Table \ref{tab:trapar} gives a summary.

\begin{table}[h]
\begin{center}
\begin{tabular}{lll}
\hline
parameter &  value \\ \hline
 $\tau$ & 0.84 to 8.4 (days) \\
 $a/r_s$ & 2 to 35 \\
 $r_p/r_s$ & 0.005 to 0.4 \\
 $i$ & 85 to 90 (degree) \\
 $u_1$ & 0.210 to 0.731 \\
 $u_2$ & 0.035 to 0.442 \\
\hline
\end{tabular}
\caption[Table R]{Transit parameters.}
\label{tab:trapar}
\end{center}
\end{table}

With the above transit model, we can have an equal number of transit 
and non-transit light curves to be used as the input of deep learning models.
For the transit ones,
a light curve of ten transit periods is employed as a standard input.
Our transit period is set to be less than 8.4 days. Thus, ten times
of it is less than 84 days, which is consistent with the maximum length
of Kepler light curves in each quarter.

The input data is the flux value of each point in interpolated light curves. 
We use 4000 points as an input in this work. 
The transit periods are known to us. We use 
400 data points for each period and include ten periods of transit.

\section{The Results}
\label{sec:compa}

\subsection{Training, Validation, and Testing Processes}
\label{ssec:laca}
We split data into three parts, 
training data, validation data, and testing data. 
The training data is used to train deep learning models.
During the training process, the parameters in models 
will be updated continuously through SGD. 
We do not input all training data at one time, 
but input small subsets one by one until all are used. 
Thus, data is separated into subsets, called mini-batches. 
We randomly assign 64 light curves to each mini-batch 
(except the final mini-batch which has less light curves). 
When all training data have been used, it is called one epoch.
Before next epoch starts, the light curves in mini-batches are 
randomly assigned again. 
There could be many epochs, and every time right after 
one epoch is completed, the model is used to predict the answers
of the validation data. The percentage of the correct answers is called
``accuracy'' in this paper. Thus, there is a value of 
accuracy for each epoch and it leads to a stopping 
condition of the training process.
After $n$ epochs are completed, if
all the accuracy of 
($n+1$)th, ($n+2$)th, ($n+3$)th, ($n+4$)th, ($n+5$)th
epochs are not larger than the accuracy of $n$th epoch, 
the training process stops at the ($n+5$)th epoch.
Typically, the number of epochs range from 10 to 30, 
depending on how quick the learning process converges. 
After that, the deep learning model is settled. It would be used
to predict the answers of testing data and
the corresponding accuracy would be calculated. 

For a given total number of light curves, 80\% are used as the training data, 
10\% are validation data, and the rest 10\% are testing data. 
For all the above three sets of data, 
a half of light curves are transit samples, 
and another half is non-transit samples. 
It is likely that the more total number of light curves we could have,
the higher accuracy it could be. However, we would need 
more computational time for the training process when there are more 
light curves. 

\begin{figure}[ht]
\centering
\includegraphics[width=1.0\textwidth]{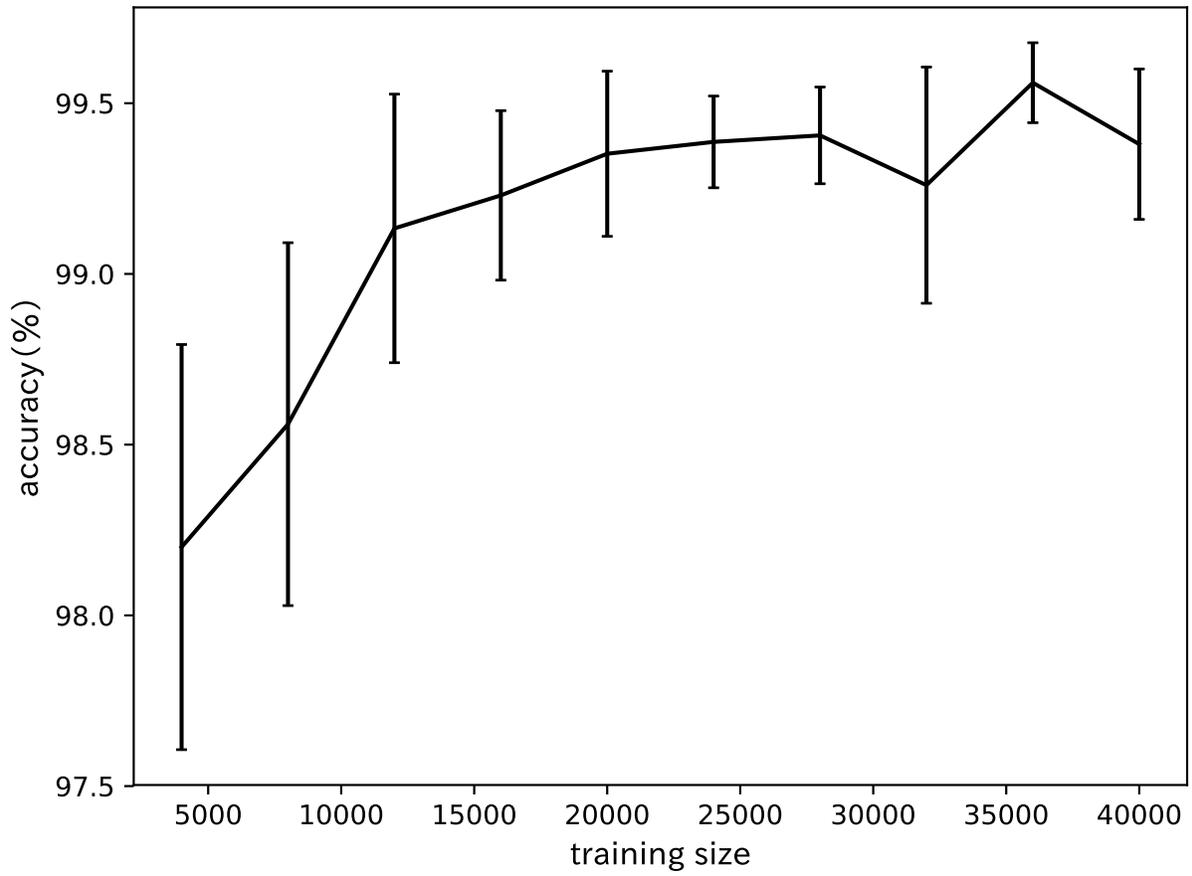}
\caption{The learning curve, i.e. the accuracy as a function of 
the training size (the number of light curves of training data). 
Model 5, i.e. 2D-CNN-folding-2, is used for this result.} 
\label{fig:laca}
\end{figure}

The learning curve, i.e.
the accuracy as a function of the number of training light curves,
of Model 5 is presented in Fig. \ref{fig:laca}.
This accuracy is calculated based on testing data after the training process 
has been done. This whole process is repeated for five times 
with different sets of validation and testing data (Stone 1974). 
The median values
are used as the results and the standard deviations give the error bars
(the results and error bars in later figures are obtained in the same way). 
In general, the accuracy stop increasing when the 
the number of training light curves is between 20000 and 40000.
Therefore, in this work, we set the total number of sample light
curves to be 50000; and among these 50000 light curves, 40000 
are chosen to be the training light curves.  

For the rest of this section, the performances
of our models will be presented. 
In addition to the accuracy, the precision (also called reliability)
and the recall (also called completeness) are also shown.
The precision is the percentage of real transit light curves
among those light curves which are predicted to have transits.
The recall is the percentage of the light curves that
are successfully predicted to have transits among 
those transit light curves. 

\subsection{Signal-to-Noise Ratios}

In order to understand how the signal-to-noise ratio (SNR) affects
the performances of all five models, 
the training, validation, and testing processes are now 
run with light curves with exactly the same SNR. 
Thus, the transit models are set through the process in Section 4
except the SNR is the same for all transit signals in one run
and $r_p/r_s$ is determined by other parameters.
The process is then repeated with different values of SNR for many runs.

Fig. 9 presents the results in which the accuracy is expressed 
as a function of SNR. 
The accuracy of Model 3-5, i.e. CNN models with folding  
are all very close to 100\% since the folding can make models 
recognize transit signals even when the SNR is low. 
When SNR is less than 10, their accuracy is still above 98\%.
The precision and recall are also shown similarly.
This confirms that the folding process is very important, 
so we shall use models with folding.

\subsection{Transit Phase Positions}
\label{ssec:feva}
For convenience, the transit signals can always be 
placed at the centers of input light curves for a deep learning model
if the folding period is the same as transit period.  
However, when the folding period becomes different from transit period,
the transit signals would locate at different positions 
of the folded light curves. In order to study this effect 
of transit phase positions, 
we define the center 
as the phase position $P_p=0.5$, the phase range parameter $P_x\in [0,1]$,  
and the phase range to be $[0.5-P_x/2, 0.5+P_x/2]$.
When $P_x=0$, transit signals are always at the center $P_p=0.5$;
when $P_x=1$, the phase range is [0,1], so transit signals are 
randomly located at any positions of folded light curves;
when $0< P_x < 1$, transit signals are uniformly  
distributed within a phase interval centered on $P_p=0.5$ with 
a width $P_x$.

For a given value of $P_x$, the transit signals are 
set to be within the phase range $[0.5-P_x/2, 0.5+P_x/2]$
randomly for each training, validation, and testing 
input light curve data.
Fig. \ref{fig:feva} shows the accuracy as a function of 
the phase range parameter $P_x$ for all five deep learning models.
It shows that the performances of Model 3-5, i.e. CNN models with folding,  
are all great
no matter what value of the phase range parameter $P_x$ is.
Their accuracy, precision, and recall are close to 100\%.      
For Model 2, i.e. 1D-CNN-folding-0, 
the accuracy drops a bit, but still about 99\%.
Due to no convolution, the accuracy of Model 1, i.e. MLP model, drops 
further to be around 93\% when the phase range parameter $P_x=1$.
Except this subsection, we always set $P_x=1$ in this paper.

\subsection{Folding Periods}
\label{ssec:pile}

When a CNN-folding model is employed to search for planetary transits,
one needs to set the folding period, which could be different from 
the unknown transit period. Thus, we shall investigate
how this difference affects the results.
As shown in Fig. \ref{fig:fefo2}, the accuracy of Model 3, i.e. 
1D-CNN-folding-1 model, will definitely decrease when the folding 
period is different from the transit period.
In order to study this effect, we define
a period difference percentage $D_f$,
and set folding periods to be 
uniformly distributed in $[(1-D_f\%)\tau, (1+D_f\%)\tau]$
for a given transit period $\tau$.
 
Fig. \ref{fig:pileer} is the result 
that folding period and transit period are 
set to be the same during the training and validation processes, 
but become different 
when the learned model is used to predict the testing data.
The accuracy is expressed as a function of the period
difference percentage $D_f$, which is set to be within [0,2].
The accuracy of Model 2, i.e. 1D-CNN-folding-0 model, 
does not decrease when $D_f$ becomes larger
because it is not a folding model.
The accuracy of other models decrease with $D_f$. 
For example, the accuracy of Model 5, i.e. 2D-CNN-folding-2 model,  
is about 85\% when $D_f=2$.

Fig. \ref{fig:piledi} is the result 
that folding period and transit period are
already different during the training and validation processes.
The models have learned these period differences.
The accuracy on testing data is plotted as a function of 
$D_f$, which is set to be within [0,20].
In this case, we found that both the accuracy of 
Model 1 (MLP model) and Model 3 (1D-CNN-folding-1 model)
drops to be about 70\% at $D_f=7.5$.  
In contrast, our Model 5, i.e. 2D-CNN-folding-2 model,
still has an accuracy about 95\% at $D_f=20$.

\section{The Demonstration}
\label{sec:demo}

From the results in Section 5, we find that 
the 2D-CNN-folding-2 model has a very good performance.
In this section, we demonstrate how to use
it to detect new transit signals.
In addition, a comparison with Christiansen et al. (2016)
will be given.

In Section 5, there are 50000 light curves (a half of them 
have injected transits), and 40000 are used as the training data,
5000 are used as the validation data, and another 5000 are 
the testing data. However, in this section, 40000 of the above 
light curves are used as the training data, and all the rest
10000 are used as the validation data.
During the above training and validation processes, 
the input of our 2D-CNN-folding-2 model is folded with 
a period difference percentage $D_f=1$, so the folding periods
are uniformly distributed in $[0.99\tau, 1.01\tau]$  
for a given transit period $\tau$ (see Section 5.4).

In order to compare with the results in Christiansen et al. (2016),
we inject transit signals into another 2000 light curves
and use them as the data with unknown transits to be searched 
and detected. 
To generate these 2000 injected transit signals, we follow
the process in Section 4 except that now $r_p/r_s$ is set to be uniformly 
distributed in the range [0.001, 0.08].

We now assume that we do not know whether there is any transit
in these 2000 light curves. To search for possible transits
by our trained 2D-CNN-folding-2 model, we need to try many 
different folding periods. We decide to have 120 folding periods 
and set them to cover the range of $\tau$ (see Table 2). 
Thus, the folding period $P_f = \alpha_i$, where $i=1, 2,...,120$,
and we set $\alpha_1 = 0.84$, $\alpha_{i+1} = 1.0195 \alpha_{i}$. 

Each light curve is then folded by the above 120 different folding periods, 
and analysed by our model. Fig. 13 presents the results of 
four different light curves. Our model gives an answer ``transit''
or ``no transit'' for each folding period. In Fig. 13(a),
we get transit detection for two folding periods and get  no detection 
for all other folding period. Thus, our 
2D-CNN-folding-2 model does detect the transit
and obtain the transit period with small uncertainty. 
For other three light curves in Fig. 13(b)-(d),
our model also detect transits successfully and obtain the
transit period with tiny uncertainties.
  
All 2000 transit light curves are analysed by our model,
and the percentage of recovered transits, i.e. the recall, is 
expressed as a function of ${\rm log_{10}} (r_p/r_s)$ in Fig. 14(a).
 The corresponding result obtained by the Kepler pipeline
in Christiansen et al. (2016) is shown as Fig. 14(b).
As we can see, when ${\rm log_{10}} (r_p/r_s) > -1.5$  
(i.e. $r_p/r_s > 0.03$), 
the recall is about 80\% for our model, but only about 60\%
for Christiansen et al. (2016).
However, when  ${\rm log_{10}} (r_p/r_s) < -2.0$ (i.e. $r_p/r_s < 0.01$), 
the recall is almost zero
for our model, but still can be around 20\% for 
Christiansen et al. (2016).
This is because that our model is not trained well for 
the light curves with small $r_p/r_s$.
As we state in Section 4, in this paper, we decide to ignore 
smaller planets due to the possible hidden transit signals 
caused by planets smaller than those discovered ones  
in Kepler non-transit light curves.
Our model could have a better performance than the Kepler pipeline
for the regime that our model is well trained. 

\section{Conclusions}
\label{sec:disca}

\begin{figure}[ht]
\centering
\includegraphics[width=0.5\textwidth]{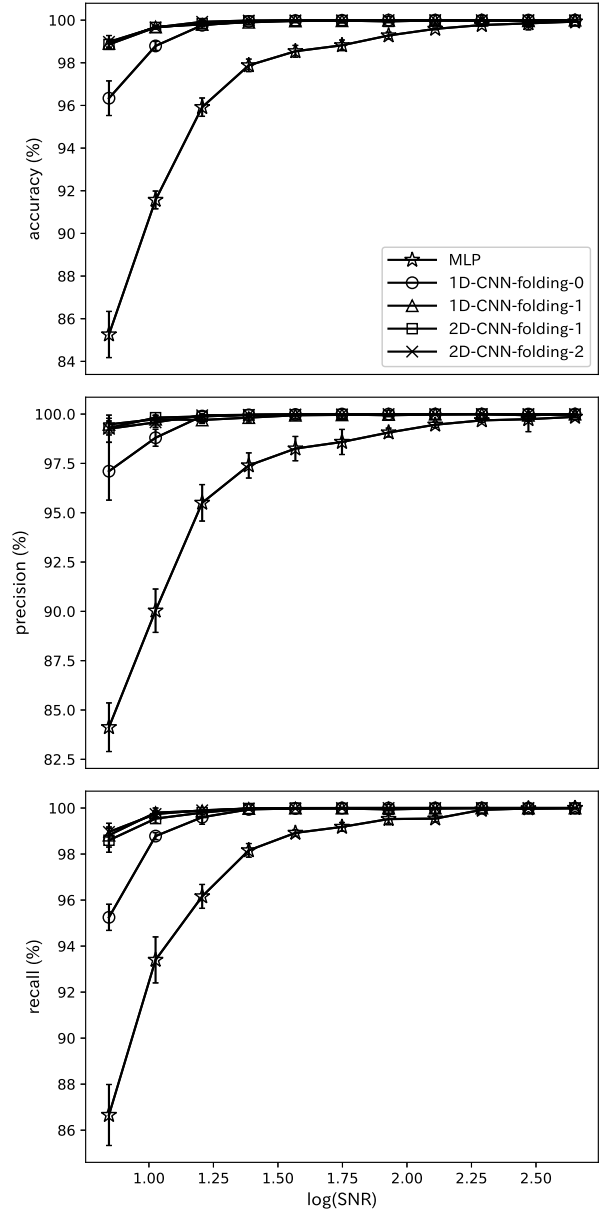}
\caption{The accuracy, precision, and recall 
as a function of SNR for all five models.
The open stars connected with solid lines are for Model 1 (MLP model);
the circles with solid lines are for Model 2 (1D-CNN-folding-0 model);
the triangles with solid lines are for Model 3 (1D-CNN-folding-1 model);
the squares with solid lines are for Model 4 (2D-CNN-folding-1 model);
the crosses with solid lines are for Model 5 (2D-CNN-folding-2 model).} 
\label{fig:accsnr}
\end{figure}

\begin{figure}[ht]
\centering
\includegraphics[width=0.5\textwidth]{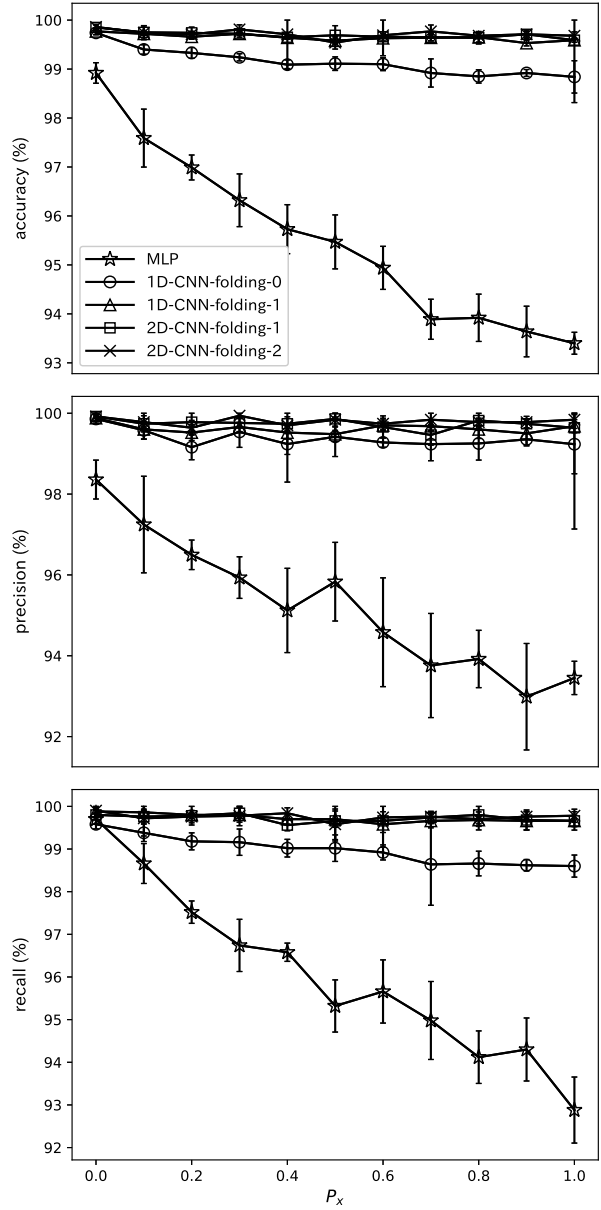}
\caption{The accuracy, precision, and recall 
as a function of phase range parameter $P_x$  
for all five models.
The open stars connected with solid lines are for Model 1 (MLP model);
the circles with solid lines are for Model 2 (1D-CNN-folding-0 model);
the triangles with solid lines are for Model 3 (1D-CNN-folding-1 model);
the squares with solid lines are for Model 4 (2D-CNN-folding-1 model);
the crosses with solid lines are for Model 5 (2D-CNN-folding-2 model).
} 
\label{fig:feva}
\end{figure}

\begin{figure}[ht]
\centering
\includegraphics[width=0.5\textwidth]{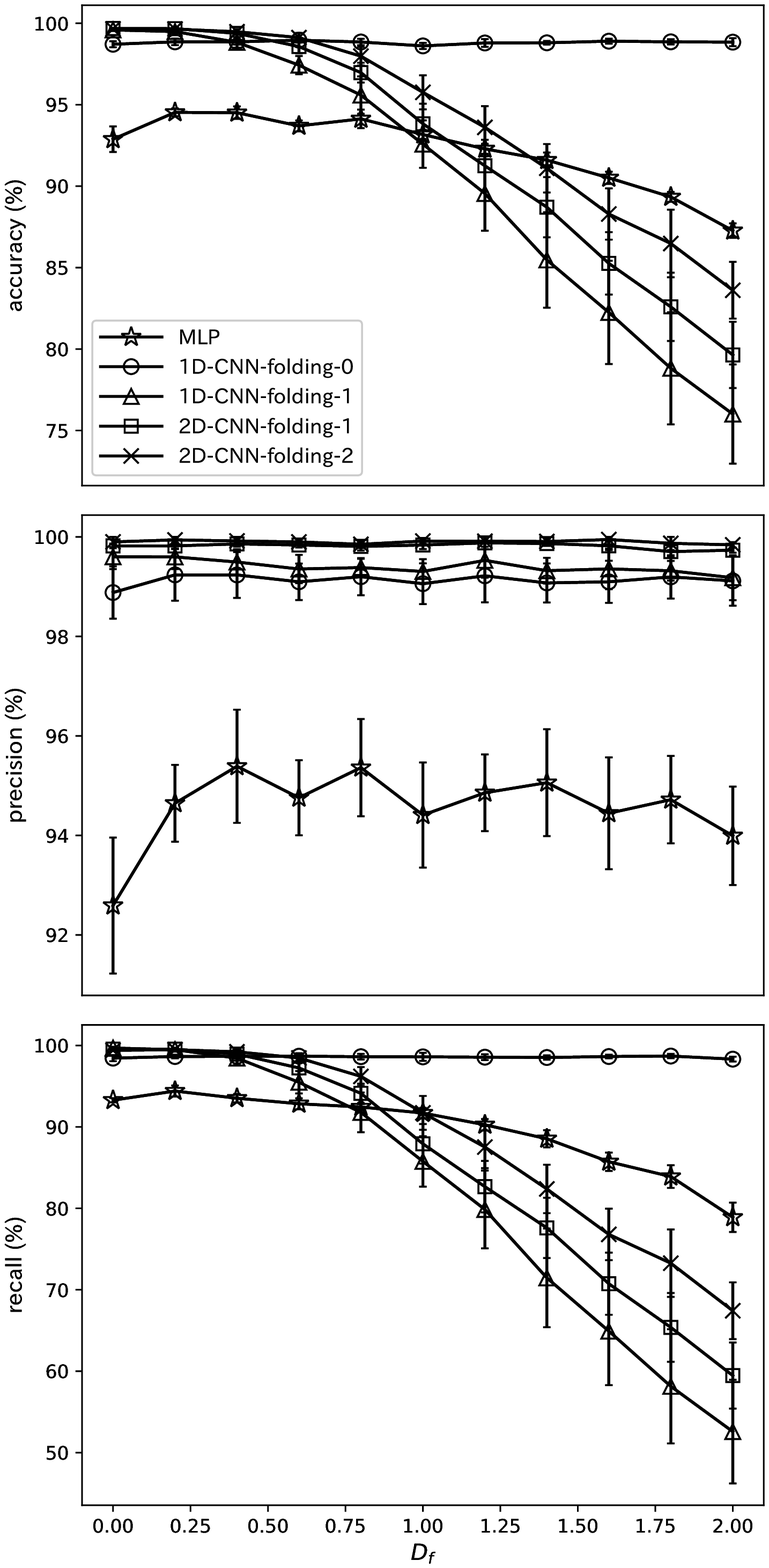}
\caption{The accuracy, precision, and recall 
as a function of period difference percentage $D_f$
for all five models. The models do not learn period differences 
during the training and validation processes.
The open stars connected with solid lines are for Model 1 (MLP model);
the circles with solid lines are for Model 2 (1D-CNN-folding-0 model);
the triangles with solid lines are for Model 3 (1D-CNN-folding-1 model);
the squares with solid lines are for Model 4 (2D-CNN-folding-1 model);
the crosses with solid lines are for Model 5 (2D-CNN-folding-2 model).
} 
\label{fig:pileer}
\end{figure}

\begin{figure}[ht]
\centering
\includegraphics[width=0.5\textwidth]{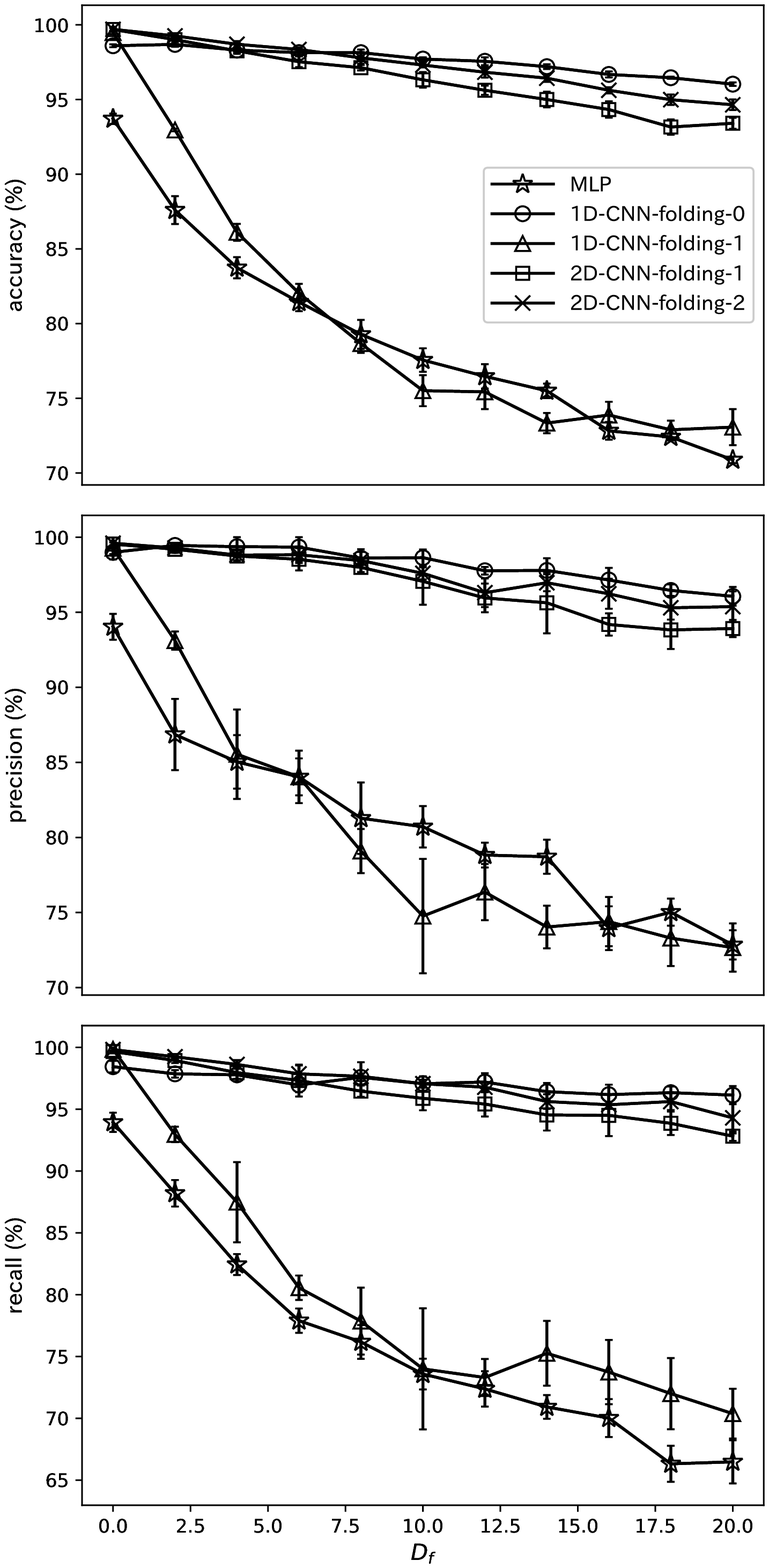}
\caption{The accuracy, precision, and recall 
as a function of period difference percentage $D_f$
for all five models. The models already learn period differences 
during the training and validation processes.
The open stars connected with solid lines are for Model 1 (MLP model);
the circles with solid lines are for Model 2 (1D-CNN-folding-0 model);
the triangles with solid lines are for Model 3 (1D-CNN-folding-1 model);
the squares with solid lines are for Model 4 (2D-CNN-folding-1 model);
the crosses with solid lines are for Model 5 (2D-CNN-folding-2 model).
} 
\label{fig:piledi}
\end{figure}

In this paper, the basic concept of machine learning was introduced.
Five deep learning models were constructed, and tested 
by the light curves from Kepler Space Telescope. 
Their performances were presented and discussed. 
From the results in Fig. \ref{fig:accsnr},
we have confirmed that the folding can help to maintain high
accuracy. All models with folding can have accuracy
above 98\% even when SNR is less than 10. 
The accuracy of models without folding
can become about 85\% when SNR is less than 10. 
The precision and recall have a similar trend. 

We have also shown that the 2D-CNN-folding-2 model can have rather good accuracy
even when the folding period is different from the transit period by 20\%.
This is an excellent feature and would be very helpful when the model
is used to search transits from new released data. 
The periods of possible transits in new data are not known, 
so we need to assume folding periods.
To show how to proceed, we have demonstrated the above process
in Section 6. We found that   
for the regime our 2D-CNN-folding-2 model is well trained, 
the transit recovered rate can be very high. 
The remained future challenge is to train the
model with transit signals caused by smaller planets.
In order to achieve that goal, we will need to generate 
more realistic non-transit light curves ourselves.
The transits by smaller planets can then be injected 
and improve the training of deep learning models.
We conclude that a deep learning model which combines
2D-CNN and folding could be an excellent tool for the future transit 
detection.

Moreover, the on-going and future space missions, such as 
Transiting Exoplanet Survey Satellite (TESS) 
and Planetary Transits and Oscillations of Stars (PLATO),
will produce huge amount of light curves.
Deep learning techniques might provide another choice
of efficient analysis tool.
   
\section*{Acknowledgment}
We are grateful to the anonymous referee for many good suggestions.
This work is supported in part 
by the Ministry of Science and Technology, Taiwan, under
Ing-Guey Jiang's Grants MOST 106-2112-M-007-006-MY3.
The data employed in this paper were obtained from 
the NASA Archives:
the Mikulski Archive for Space Telescopes (MAST) 
and the Exoplanet Archive.

\clearpage

\clearpage

\clearpage

\clearpage

\clearpage

\clearpage

\clearpage

\clearpage

\clearpage

\clearpage

\clearpage

\clearpage
\begin{figure}[ht]
\centering
\includegraphics[width=1.0\textwidth]{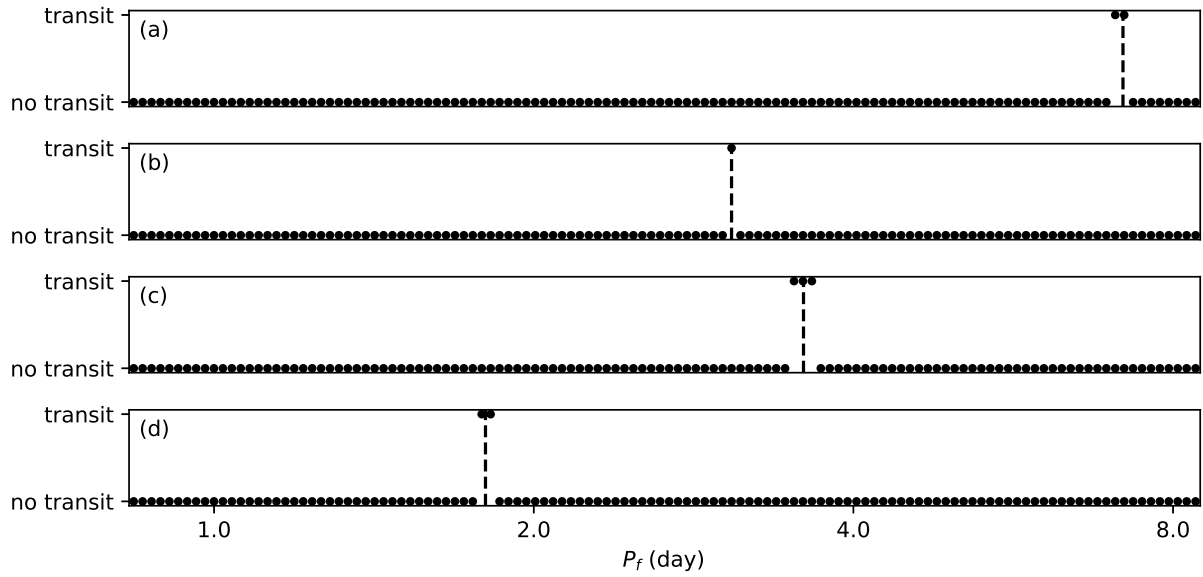}
\caption{The result of transit detection as a function of folding periods
for four light curves with injected transits.
The dashed vertical lines indicate the period of injected transits.
The full circles indicate the folding periods.} 
\label{fig13}
\end{figure}

\clearpage
\begin{figure}[ht]
\centering
\includegraphics[width=1.0\textwidth]{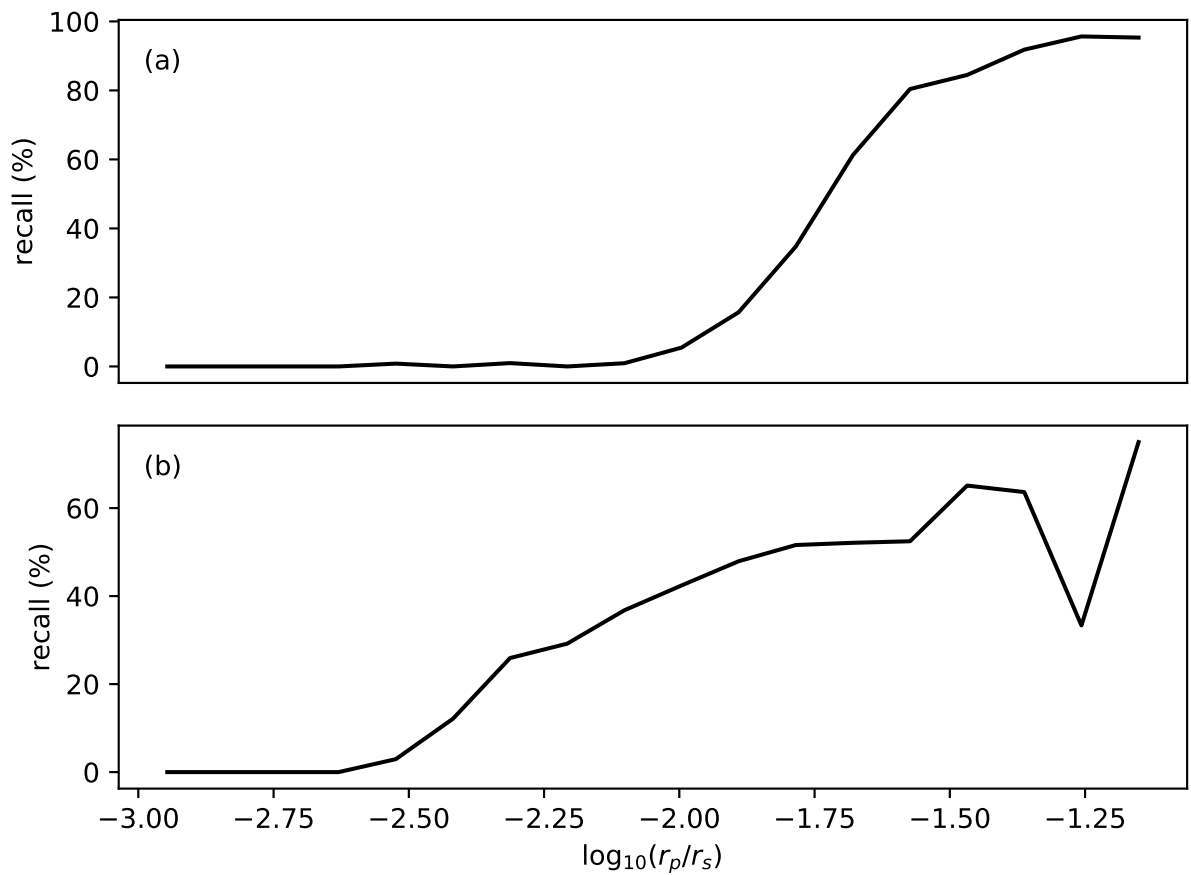}
\caption{The recall as a function of ${\rm log_{10}} (r_p/r_s)$.
Panel (a) is the result from our 2D-CNN-folding-2 model,
and Panel (b) is the result from Christiansen et al. (2016).} 
\label{fig14}
\end{figure}

\end{document}